\documentclass{article}
\title{Flavor-number dependence of QCD vacuum tensor susceptibilities}
\author{Hongting Yang \\
       {\small\sl Department of Physical Science and Technology, College of Science}\\
       {\small\sl Wuhan University of Technology, Wuhan 430070, China}}
\date{}

\begin{document}
\maketitle
\begin{abstract}
The values of quark condensate and tensor susceptibility of QCD
vacuum are calculated with different number of flavors. It is shown
that, the values of tensor susceptibility are obviously dependent on
the flavor numbers while the flavor-number dependence of quark
condensate is undetermined.

\end{abstract}
{\it E-mail}\/: yht@mail.whut.edu.cn
\newpage

The quark condensate and tensor susceptibility are two quantities of
reflecting the nonperturbative aspects of QCD vacuum. The recent
calculations show that, there exists a great discrepancy between our
values of the QCD vacuum tensor susceptibilities and those obtained
from QCD sum rules and from chiral constituent quark
model~\cite{HH03}. Within the framework of chiral perturbation
theory~\cite{JH85a}, Moussallam argued that the quark condensate is
strongly dependent on the flavor-number $N_f$  via the calculation
of the chiral coupling-constant $L_6$~\cite{B00}. This strong $N_f$
dependence cannot be derived in our calculations below, while on the
other hand, we find the strong $N_f$ dependence of the values of
tensor susceptibilities. Our calculations are performed within the
framework of an effective QCD model, the global color symmetry model
(GCM)~\cite{RC85}, which has been well developed in the last twenty
years~\cite{HJ03}.

The GCM partition function for massless quarks in Euclidean space is
written as
\begin{equation}
   {\cal Z}_{\rm GCM}=\int{\cal D}\bar{q}{\cal D}q{\cal D}Ae^{-S_{\rm GCM}[\bar{q},q,A]}
\end{equation}
with the GCM action
\begin{equation}
   S_{\rm GCM}[\bar{q},q,A]=\int{\rm d}x[\bar{q}(x)({\not\!\partial}-ig{\not\!\!A})q(x)+
   \int{\rm d}x{\rm d}y{1\over 2}A_{\mu}^a(x)[D^{ab}_{\mu\nu}(x-y)]^{-1}A_{\nu}^b(y).
\end{equation}
Here $D^{ab}_{\mu\nu}(x-y)$ is the gluon 2-point Green's function,
and a Feynman-like gauge
$D^{ab}_{\mu\nu}(x-y)=\delta_{\mu\nu}\delta^{ab}D(x-y)$ is used. The
important dynamical characteristics of local color symmetry are
included in $D(x-y)$. The phenomenological gluon propagator is
treated as the model input parameter and is adjusted to reproduce
the pion decay constant.

Applying the standard bosonization procedure, the resulting
expression for the GCM partition function in terms of the bilocal
field is given by
\begin{equation}
  {\cal Z}_{\rm GCM}=\int{\cal D}{\cal B}^{\theta}\exp\left(-S[{\cal B}^{\theta}]\right),
\end{equation}
where the action is
\begin{equation}
   S[{\cal B}^{\theta}]=-{\rm Tr\,Ln}[G^{-1}]+\int{\rm d}x{\rm d}y
   \frac{{\cal B}^{\theta}(x,y){\cal B}^{\theta}(y,x)}{2g^2D(x-y)}\,.
\end{equation}
The quark inverse Green's function $G^{-1}$ is defined as
\begin{equation}
   G^{-1}(x,y)={\not\!\partial}\delta(x-y)+\Lambda^{\theta}{\cal B}^{\theta}(x,y)\,,
\end{equation}
where the quantity $\Lambda^{\theta}={1\over2}K^aC^bF^c$ arises from Fierz reordering of the
current-current interaction
\begin{equation}
 \Lambda^{\theta}_{ji}\Lambda^{\theta}_{lk}=(\gamma_{\mu}\frac{\lambda^a}{2})_{jk}
 (\gamma_{\mu}\frac{\lambda^a}{2})_{li}\,.
\end{equation}
Here the Dirac matrix is
$K^a=(I_D,i\gamma_5,\frac{i}{\sqrt{2}}\gamma_{\mu},\frac{i}{\sqrt{2}}\gamma_{\mu}\gamma_5)$,
the color matrix is $C^b=({4\over 3}I_c,
\frac{i}{\sqrt{3}}\lambda^a_c)$, and the flavor matrices for
$N_F=2$ flavors and $N_F=3$ flavors are respectively
$F^c_{N_F=2}=({1\over\sqrt{2}}I_F,
{1\over\sqrt{2}}{\vec{\tau}_F})$ and
$F^c_{N_F=3}=({1\over\sqrt{3}}I_F,
{1\over\sqrt{2}}{\lambda_F^a})$.

The saddle point of the action is obtained by minimizing the
bilocal action: $\left. \frac{\delta S[{\cal B}]}{\delta {\cal B}}
\right |_{{\cal B}_0}=0,$ which gives
\begin{equation}
  {\cal B}^{\theta}_0(x)=g^2D(x){\rm tr}[G_0(x)\Lambda^{\theta}]\,.
\end{equation}
We define $\Sigma(p)\equiv\Lambda^{\theta}{\cal B}^{\theta}_0(p)
=i{\not\!p}[A(p^2)-1]+B(p^2)$, where the self-energy functions $A$
and $B$ satisfy the Dyson-Schwinger equations,
\begin{eqnarray}
   [A(p^2)-1]p^2 &=&{8\over 3}\int\frac{{\rm d}^4q}{(2\pi)^4}\ g^2D\left((p-q)^2\right)\
   \frac{A(q^2)q\cdot p}{q^2A^2(q^2)+B^2(q^2)}\,,  \label{DSE1}  \\ \label{DSE2}
   B(p^2) &=&{16\over 3}\int\frac{{\rm d}^4q}{(2\pi)^4}\ g^2D\left((p-q)^2\right)\
   \frac{B(q^2)}{q^2A^2(q^2)+B^2(q^2)}\,.
\end{eqnarray}
For the convenience of computation, the Dyson-Schwinger equations
are often rewritten as
\begin{eqnarray}
   [A(s)-1]s &=& \frac{1}{3\pi^3}\int_0^\infty s^\prime{\rm d}s^\prime\int_0^\pi{\sin^2\chi}
   g^2D(s,s^\prime)\frac{\sqrt{ss^\prime}A(s^\prime)\cos\chi}{s^\prime A^2
   (s^\prime)+B^2(s^\prime)}{\rm d}\chi\,,   \\
   B(s) &=& \frac{2}{3\pi^3}\int_0^\infty s^\prime{\rm d}s^\prime\int_0^\pi{\sin^2\chi}
   g^2D(s,s^\prime)\frac{B(s^\prime)}{s^\prime A^2
   (s^\prime)+B^2(s^\prime)}{\rm d}\chi\,,
\end{eqnarray}
where $s=p^2$, and
$g^2D(s,s^\prime)=g^2D(s+s^\prime-2\sqrt{ss^\prime}\cos\chi)$.

The quark Green's function at ${\cal B}^{\theta}_0$ is given by
\begin{equation}
  G_0(x,y)=G_0(x-y)=\int\frac{{\rm d}^4p}{(2\pi)^4}
  \frac{-i{\not\!p}A(p^2)+B(p^2)}{p^2A^2(p^2)+B^2(p^2)}e^{ip\cdot(x-y)}\,,
\end{equation}
and the equation used to calculate $f_\pi$ is
\begin{equation}  \label{fpi}
   {f_\pi}^2=\frac{3N_F}{16\pi^2}\int_0^\infty s{\rm d}s\ \frac{A^2(s)B^2(s)}{[sA^2(s)+B^2(s)]^2}
   \left[2+\frac{B^2(s)}{sA^2(s)+B^2(s)}\right]\,.
\end{equation}
In Eq.~(\ref{fpi}), all those terms involving the derivatives of
$A(s)$ and $B(s)$ with respect to $s$ are neglected. Let $A(s)=1$,
$B(s)=M$, with $M$ the mass of the constituent quark, the integrand
of Eq.~(\ref{fpi}) behaves like $2M^2/s$ when $s$ approaches
infinity, which reproduce the result of Ref.~\cite{WM98} strictly.

The two quantities to be evaluated are quark condensate
\begin{equation}
  \langle\bar{q}q\rangle=-\frac{3}{4\pi^2}\int^\infty_0s{\rm d}s\ \frac{B(s)}{sA^2(s)+B^2(s)}\,,
\end{equation}
and the quantity \(\Pi_{\chi}(0)\) defined as
\begin{equation}
  {1\over 12}\Pi_{\chi}(0)\equiv-\frac{3}{4\pi^2}
  \int^\infty_0s{\rm d}s\left [\frac{B(s)}{sA^2(s)+B^2(s)}\right ]^2\,,
\end{equation}
which relates the quark condensate to the tensor susceptibility $\chi$ via the definition
\begin{equation}  \label{chi}
  \chi\equiv\frac{\Pi_{\chi}(0)}{6\langle\bar{q}q\rangle}\,.
\end{equation}

We choose three different gluon propagators as follows, the ultraviolet behavior of them are
different from that in QCD~\cite{PC97,KH99}. These models are numbered as model 1:
\begin{equation}
   g^2D(s)=3\pi^2\frac{\lambda^2}{\Delta^2}e^{-{s\over\Delta}}\,,
\end{equation}
 model 2:
\begin{equation}
g^2D(s)=3\pi^2\frac{\lambda^2}{\Delta^2}e^{-{s\over\Delta}}+\frac{4\pi^2d}{s\ln[s/\Lambda^2+e]}\,,
\end{equation}
and model 3:
\begin{equation}
g^2D(s)=4\pi^2d\frac{\lambda^2}{s^2+\Delta}\,.
\end{equation}
Here $d=12/(33-2N_f)$ and $\Lambda=200$ MeV.

The calculation results for model 1 are displayed in
Table~\ref{Table1}, and those for model 2 and model 3 are
respectively given in Tables~\ref{Table2} and \ref{Table3}.
\begin{table}[htb]
\begin{center}
\caption{\label{Table1} $\langle\bar{q}q\rangle$ and $\Pi_{\chi}(0)$
are calculated for model 1:
$g^2D(s)=3\pi^2\frac{\lambda^2}{\Delta^2}e^{-{s\over\Delta}}$. The
parameters $\Delta$ and $\lambda$ are adjusted to fit
$f_\pi=87$~MeV.}
\begin{tabular}{ccccccccc}
\hline \multicolumn{4}{c}{$N_F=2$} & & \multicolumn{4}{c}{$N_F=3$}
\\ \cline{1-4} \cline{6-9} $\Delta$ & $\lambda$ &
$-\langle\bar{q}q\rangle^{1/3}$ & $-\Pi_{\chi}(0)/12$ & & $\Delta$
&$\lambda$& $-\langle\bar{q}q\rangle^{1/3}$ & $-\Pi_{\chi}(0)/12$ \\
$[{\rm GeV}^2]$ & $[{\rm GeV}]$ & $[{\rm MeV}]$ & $[\times10^3\ {\rm
MeV}^2]$ & & $[{\rm GeV}^2]$ & $[{\rm GeV}]$ & $[{\rm MeV}]$ &
$[\times10^3\ {\rm MeV}^2]$
\\ \hline
 0.200& 1.509 & 207 & 1.71 & & 0.200& 1.309 & 180 & 1.25  \\
 0.300& 1.604 & 221 & 1.87 & & 0.300& 1.439 & 194 & 1.35  \\
 0.400& 1.709 & 233 & 1.98 & & 0.400& 1.568 & 206 & 1.42  \\
 0.500& 1.816 & 243 & 2.07 & & 0.500& 1.690 & 216 & 1.46  \\ \hline
\end{tabular}
\end{center}
\begin{center}
\caption{\label{Table2} Same as Table~\ref{Table1} for model 2:
$g^2D(s)=3\pi^2\frac{\lambda^2}{\Delta^2}e^{-{s\over\Delta}}+
\frac{4\pi^2d}{s\ln[s/\Lambda^2+e]}$,
$d=12/(33-2N_f)$, $\Lambda=200$ MeV.}
\begin{tabular}{ccccccccc}
\hline \multicolumn{4}{c}{$N_F=2$} & & \multicolumn{4}{c}{$N_F=3$}
\\ \cline{1-4} \cline{6-9} $\Delta$ & $\lambda$ &
$-\langle\bar{q}q\rangle^{1/3}$ & $-\Pi_{\chi}(0)/12$ & & $\Delta$
&$\lambda$& $-\langle\bar{q}q\rangle^{1/3}$ & $-\Pi_{\chi}(0)/12$ \\
$[{\rm GeV}^2]$ & $[{\rm GeV}]$ & $[{\rm MeV}]$ & $[\times10^3\ {\rm
MeV}^2]$ & & $[{\rm GeV}^2]$ & $[{\rm GeV}]$ & $[{\rm MeV}]$ &
$[\times10^3\ {\rm MeV}^2]$ \\ \hline
 0.100& 1.453 & 216 & 1.33 & & 0.100& 1.213 & 188 & 0.925  \\
 0.200& 1.509 & 232 & 1.52 & & 0.200& 1.306 & 205 & 1.07  \\
 0.300& 1.593 & 245 & 1.66 & & 0.300& 1.420 & 218 & 1.16  \\
 0.400& 1.687 & 256 & 1.76 & & 0.400& 1.536 & 228 & 1.22  \\ \hline
\end{tabular}
\end{center}
\begin{center}
\caption{\label{Table3} Same as Table~\ref{Table1} for model 3:
$g^2D(s)=4\pi^2d\frac{\lambda^2}{s^2+\Delta}$, $d=12/(33-2N_f)$.}
\begin{tabular}{ccccccccc}
\hline \multicolumn{4}{c}{$N_F=2$} & & \multicolumn{4}{c}{$N_F=3$}
\\ \cline{1-4} \cline{6-9} $\Delta$ & $\lambda$ &
$-\langle\bar{q}q\rangle^{1/3}$ & $-\Pi_{\chi}(0)/12$ & & $\Delta$
&$\lambda$& $-\langle\bar{q}q\rangle^{1/3}$ & $-\Pi_{\chi}(0)/12$ \\
$[{\rm GeV}^4]$ & $[{\rm GeV}]$ & $[{\rm MeV}]$ & $[\times10^3\ {\rm
MeV}^2]$ & & $[{\rm GeV}^4]$ & $[{\rm GeV}]$ & $[{\rm MeV}]$ &
$[\times10^3\ {\rm MeV}^2]$ \\ \hline
 $10^{-2}$& 1.345 & 243 & 1.47 & & $10^{-2}$& 1.157 & 207 & 1.03  \\
 $10^{-3}$& 1.109 & 223 & 1.34 & & $10^{-3}$& 0.930 & 187 & 0.914  \\
 $10^{-4}$& 0.955 & 210 & 1.29 & & $10^{-4}$& 0.791 & 175 & 0.864  \\
 $10^{-5}$& 0.847 & 201 & 1.28 & & $10^{-5}$& 0.695 & 167 & 0.850  \\ \hline
\end{tabular}
\end{center}
\end{table}
It is indicated that, the value of $f_\pi$ in 2-flavor QCD is
different from the one in 3-flavor QCD~\cite{JH85b}. However, this
difference is so small that almost we need not make any
modifications to the calculation results. In the model 3 for
example, when $N_f=2$, if we take $\Delta=10^{-2}\ {\rm GeV}^4$,
$\lambda=1.355\ {\rm GeV}$, $f_\pi$ is fitted to 88 MeV, the values
of $-\langle\bar{q}q\rangle^{1/3}$ and $-\Pi_{\chi}(0)/12$ are
respectively 245 MeV and $1.51\times10^3\ {\rm MeV}^2$, which are
very close to the corresponding data 243 MeV and $1.47\times10^3\
{\rm MeV}^2$ for $f_\pi=87$ MeV. So we take $f_\pi=87\ {\rm MeV}$ in
all cases.

Let's analyse the results listed in Table~\ref{Table1} for model 1.
If $\Delta$ is independent on $N_f$, say, $\Delta=0.200\ {\rm
GeV}^2$, then the parameter $\lambda=1.509\ {\rm GeV}$ for $N_f=2$
should be replaced by $\lambda=1.309\ {\rm GeV}$ for $N_f=3$ to fit
the pion decay constant, which means the parameter $\lambda$ is
dependent on $N_f$. Similarly, if $\lambda$ is independent on $N_f$,
then $\Delta$ must be dependent on $N_f$ (the calculations for this
case are not given here for clarity). So, in any case, there is at
least one parameter which is dependent on $N_f$. Obviously, if
$\Delta$ is independent on $N_f$, the values of quark condensate
$-\langle\bar{q}q\rangle^{1/3}$ are universally decreased by more
than $10\%$ when $N_f$ changes from 2 to 3. However, the $N_f$
dependence of $\Delta$ in GCM is actually unknown. On the other
hand, when $\Delta$ moves from $0.200\ {\rm GeV}^2$ to $0.500\ {\rm
GeV}^2$, the quark condensate $-\langle\bar{q}q\rangle^{1/3}$ for
$N_f=2$ ranges from 207 MeV to 243 MeV. If $\Delta<0.200\ {\rm
GeV}^2$ or $\Delta>0.500\ {\rm GeV}^2$, the corresponding values of
quark condensate will leave this range. The values of quark
condensate for $N_f=3$ ranges from 180 MeV to 216 MeV, which partly
coincides with the one for $N_f=2$. Consequently the $N_f$
dependence of quark condensate is unsure. In contrast to this, the
$N_f$ dependence of tensor susceptibilities is evident. The values
of $-\Pi_{\chi}(0)/12$ for $N_f=2$ ranges from $1.71\times10^3\ {\rm
MeV}^2$ to $2.07\times10^3\ {\rm MeV}^2$, while that for $N_f=3$
decrease to $(1.25\sim1.46)\times10^3\ {\rm MeV}^2$.

It is easy to see that, the similar peculiarities for the values of
$-\langle\bar{q}q\rangle^{1/3}$ and $-\Pi_{\chi}(0)/12$ can also be
found both in Tables~\ref{Table2} and \ref{Table3}.

In conclusion, for three models of gluon propagators, the values of
$-\langle\bar{q}q\rangle^{1/3}$ and $-\Pi_{\chi}(0)/12$ are obtained
by adjusting the parameters $\lambda$ and $\Delta$ to fit the pion
decay constant. Because the accurate flavor-number dependence of the
gluon propagators (i.e. the $N_f$ dependence of $\lambda$ and
$\Delta$) is unclear, the flavor-number dependence of two quark
condensate is undetermined. Compared with the results for $N_F=2$
flavors, the values of $-\Pi_{\chi}(0)/12$ for $N_F=3$ flavors are
universally lowered. From Eq.~\ref{chi}, the definition of tensor
susceptibility, for an average value of quark condensate, a smaller
value of $-\Pi_{\chi}(0)/12$ implies a smaller value of tensor
susceptibility. The tensor susceptibilities of QCD vacuum are
relevant with the flavor numbers.

\end{document}